\documentclass[twocolumn,           % Format:   preprint, twocolumn
               showpacs,            % Pacs  :   showpacs, noshowpacs
               preprintnumbers,     % Preprint: preprintnumbers, 	
               aps,                 % Society:
               prd,          	    % Style : pra,prb,prc,prd,pre,prl,
               a4paper,             % Size  :   a4paper, ...
               superscriptaddress,  % Title :   groupedaddress, 
               unsortedaddress,
               footinbib,           % Foonote:  footinbib, nofootinbib 
               tightenlines,        % Remove additional spaces in a line
               floats               % Floating pictures and tables
               ]{revtex4}

\usepackage{graphicx}
\usepackage{amssymb,latexsym}
\usepackage[draft=false]{hyperref}

\newcommand{\ba}{\begin{eqnarray}}
\newcommand{\ea}{\end{eqnarray}}  
\newcommand{\be}{\begin{equation}}
\newcommand{\ee}{\end{equation}}

\begin{document}

\hyphenation{brane-world}  

%%--- DRAFTCOPY --------------------------------
%% Prints a large "DRAFT" diagonally across each page
%% Does not show up in TeXview
%% \typeout{Prints "DRAFT" on each page; does not show in TeXView}
% \special{!userdict begin /bop-hook{gsave 200 30 translate
% 65 rotate /Times-Roman findfont 216 scalefont setfont
% 0 0 moveto 0.90 setgray (DRAFT) show grestore}def end}
%%------------------------------------------------

%======================================%
%<<<<<<<<<<<< TITLE PAGE >>>>>>>>>>>>>>%
%======================================%

%\renewcommand{\topfraction}{0.99}
%\renewcommand{\bottomfraction}{0.99}

\title{Inflaton potential reconstruction in the braneworld scenario}
\author{Andrew R.~Liddle}
\affiliation{Astronomy Centre, University of Sussex, 
             Brighton BN1 9QJ, United 
Kingdom}
\author{A.~N.~Taylor}
\affiliation{Institute for Astronomy, Royal Observatory,
Blackford Hill, Edinburgh, EH9 3HJ, United Kingdom}
\date{\today} 
\pacs{98.80.Cq \hfill astro-ph/0109412}
\preprint{astro-ph/0109412}

%======================================%
%<<<<<<<<<<<<< ABSTRACT >>>>>>>>>>>>>>>%
%======================================%

\begin{abstract}
We consider inflaton potential reconstruction in the context of the simplest 
braneworld scenario, where both the Friedmann equation and the form of scalar 
and tensor perturbations are modified at high energies. We derive the 
reconstruction equations, and analyze them analytically in the high-energy limit 
and numerically for the general case. As previously shown by Huey and 
Lidsey, the consistency equation between scalar and tensor perturbations is 
unchanged in the braneworld scenario. We show that this leads to a perfect 
degeneracy in reconstruction, whereby a different viable potential can be 
obtained for any value of the brane tension $\lambda$. Accordingly, the initial
perturbations alone cannot be used to distinguish the braneworld scenario from 
the 
usual Einstein gravity case.
\end{abstract}

\maketitle
%======================================%
%<<<<<<<<<<<< MAIN TEXT  >>>>>>>>>>>>>>%
%======================================%

%%%%%%%%%%%%%%%%%%%%%%%%%%%%%%%%%%%%%%%%%%%%%%%%%%%%%%%%%%%%%%%%
\section{Introduction}

Upcoming observations, particularly of cosmic microwave anisotropies, have the 
prospect of placing the first serious constraints on models for the origin of 
structure, amongst which inflation is currently the leading candidate (see 
Ref.~\cite{LL} for an extensive review). Provided the inflationary paradigm 
remains successful, an ultimate goal is the `reconstruction' of the inflaton 
potential from observable quantities \cite{CKLL,LLKCBA}. Attention has primarily 
been focussed on the case of single-field slow-roll inflation, as it provides 
a class of simple models which can be considered within a common framework.

In order to have a robust interpretation of upcoming observations, such as those 
of cosmic microwave background anisotropies by the {\it MAP} and {\it Planck} 
satellites, it is imperative to develop an understanding of how the 
reconstruction process may be affected by degeneracies, whereby different 
cosmological models give rise to identical or near-identical observational 
predictions. Degeneracies have been considered in some depth as far as the usual 
cosmological parameters are concerned, the most important one being connected to 
the angular diameter distance to the last-scattering surface. As far as the 
initial perturbations are concerned, the most significant degeneracy discussed 
thus far is an approximate degeneracy between a tensor contribution to the 
anisotropies and the effect of early reionization, which though approximate 
requires accurate polarization measurements to be significantly broken.

A different possible degeneracy concerns the inflationary models themselves, and 
whether there might be different models giving rise to the same initial spectra 
of scalar and tensor perturbations. For single-field slow-roll models, there is 
a unique correspondence between the tensor spectrum and the inflationary 
potential, whereas if only the scalar perturbations can be observed there 
remains a one-parameter perfect degeneracy leading to a family of possible 
potentials \cite{CKLL}. In this paper, we consider whether there might be 
additional 
degeneracies associated with the possibility that the physics of inflation might 
go beyond Einstein gravity with a single scalar field. Specifically, we consider 
the simplest incarnation of the braneworld scenario, in which at high energies 
both the classical background evolution and the form of the perturbations are 
modified. We will show that this leads to a perfect degeneracy even if both 
tensor and scalar perturbations are measured. This shows that the initial 
perturbations 
alone cannot be used to distinguish the braneworld scenario from the usual 
Einstein 
gravity one.

\section{Basic formulae}

We restrict ourselves throughout to the simplest braneworld scenario 
\cite{bin,shiro}, based on the type-II Randall--Sundrum model \cite{RSII}, where 
the 
Friedmann equation receives an additional term 
quadratic in the density. We use the slow-roll approximation, as formulated by 
Maartens et al.~\cite{MWBH}. Using the definitions of Ref.~\cite{LLKCBA}, the 
spectra of scalar \cite{MWBH} and tensor \cite{LMW,HL} spectra are given by
\begin{eqnarray}
A_{{\rm S}}^2 & = & \frac{4}{25} \, \frac{H^2}{\dot{\phi}^2} \, \left( 
\frac{H}{2\pi} \right)^2 
 \simeq \frac{512 \pi}{75 M_4^6} \, \frac{V^3}{V'^2} \, \left( 1 + 
\frac{V}{2\lambda} \right)^3 \,; \\
A_{{\rm T}}^2 & = & \frac{4}{25\pi} \, \frac{H^2}{M_4^2} \, F^2(H/\mu) 
\label{tensamp} \,,
\end{eqnarray}
where
\begin{eqnarray}
F(x) & = & \left[\sqrt{1+x^2} - x^2 \ln \left( \frac{1}{x} + \sqrt{1 + 
\frac{1}{x^2}} \right) \right]^{-1/2} \,, \nonumber \\
& = & \left[\sqrt{1+x^2} - x^2 \sinh^{-1} \frac{1}{x} \right]^{-1/2} 
\label{functionF}\,,
\end{eqnarray}
and the slow-roll approximation has been used to obtain expressions in terms of 
the 
inflaton potential $V(\phi)$. Here $M_4$ is the four-dimensional Planck mass, 
and $\lambda$ is the brane tension. The mass scale $\mu$ is given by
\begin{equation}
\mu = \sqrt{\frac{4\pi}{3}} \, \sqrt{\lambda} \, \frac{1}{M_4} \,,
\end{equation}
prime indicates derivative with respect to the scalar field $\phi$, and dot a 
derivative with respect to time.

The Hubble parameter $H$ is related to the energy density $\rho$ by 
\begin{equation}
\label{Hubble}
H^2 = \frac{8\pi}{3 M_4^2} \, \rho \, \left(1 + \frac{\rho}{2\lambda} \right)
\simeq \frac{8\pi}{3 M_4^2} \, V \, \left(1 + \frac{V}{2\lambda} \right) \,,
\end{equation}
which reduces to the usual Friedmann equation for \mbox{$\rho \ll \lambda$}, the 
scalar 
field obeys the usual slow-roll equation
\begin{equation}
3H \dot{\phi} \simeq -V' \,,
\end{equation}
and the amount of expansion, in terms of $e$-foldings, is given by
\begin{equation}
\label{efolds}
\frac{dN}{d\phi} = \frac{H}{\dot{\phi}}
\end{equation}

The expressions for the spectra are, as always, to be evaluated at Hubble radius 
crossing $k = aH$, and the spectral indices of the scalars and tensors are 
defined as usual by
\begin{equation}
n-1 \equiv \frac{d \ln A_{{\rm S}}^2}{d \ln k} \quad ; \quad n_{{\rm T}} \equiv 
\frac{d \ln A_{{\rm T}}^2}{d \ln k} \,.
\end{equation}
If one defines slow-roll parameters, generalizing the usual ones, by \cite{MWBH}
\begin{eqnarray}
\epsilon_{{\rm B}} & \equiv & \frac{M_4^2}{16\pi} \, \left( 
	\frac{V'}{V} \right)^2 \; \frac{1 + V/\lambda}{\left(1 +
	V/2\lambda \right)^2} \,; \\
\eta_{{\rm B}} & \equiv & \frac{M_4^2}{8\pi} \, \frac{V''}{V} \; 
	\frac{1}{1+V/2\lambda}  \,,
\end{eqnarray}
then the scalar spectral index, in the slow-roll approximation, obeys the usual 
equation
\begin{equation}
n - 1 \simeq -6 \epsilon_{{\rm B}} + 2 \eta_{{\rm B}} \,.
\end{equation}

The tensor index obeys a more complicated equation, but remarkably it still 
obeys the usual consistency equation \cite{HL} 
\begin{equation}
R \equiv \frac{A_{{\rm T}}^2}{A_{{\rm S}}^2} = -\frac{1}{2} n_{{\rm T}} \,.
\end{equation}
This result is maintained because the normalization of the tensor mode function 
requires $F(x)$ to obey a particular differential equation.
This tells us that the relation between scalar and tensor perturbations is 
unchanged in the braneworld scenario. In particular, this equation can be viewed 
as an expression giving the scalar spectrum corresponding to a given tensor 
spectrum, namely 
\begin{equation}
A_{{\rm S}}^2 = -2\frac{A_{{\rm T}}^2}{n_{{\rm T}}} = - \frac{A_{{\rm 
T}}^3}{dA_{{\rm T}}/d \ln k} \,.
\end{equation}
Hence in the braneworld scenario, as in Einstein gravity, the scalars carry no 
additional information about the potential if the tensors are known. This 
invariance of the consistency relation under a change in gravitational physics 
is 
unexpected, since in most variations of standard inflation, e.g.~warm inflation 
\cite{TB00}, this relation is substantially different.

\subsection{Low-energy limit}

Provided $\rho \ll \lambda$, the equations all reduce to the usual Einstein 
gravity ones and the normal reconstruction equations apply. In particular, if 
one is able to observe the tensor spectrum, it gives a unique potential (under 
the single-field assumption), and if the scalars can additionally be measured 
they obey consistency relations. If only the scalars 
can be measured, a unique potential cannot be obtained and there is a 
one-parameter family of possible models. Measurement of the tensors at a single 
scale is sufficient to remove this degeneracy.

\subsection{High-energy limit}

Before proceeding to the general case, it is instructive to analyze the 
high-energy limit. The key expressions are
\begin{eqnarray}
A_{{\rm S}}^2 & \simeq & \frac{64\pi}{75 M_4^6} \, \frac{V^6}{V'^2 \lambda^3} 
\,; \\
A_{{\rm T}}^2 & \simeq & \frac{8}{25 M_4^4} \, \frac{V^3}{\lambda^2} \,; \\
n-1 & \simeq & - \frac{M_4^2}{4\pi} \, \frac{\lambda}{V} \left[ 6 
\left(\frac{V'}{V}\right)^2 - 2 \frac{V''}{V}\right] \,; \\
n_{{\rm T}} & \simeq & - \frac{3M_4^2}{4\pi} \, \frac{\lambda}{V} \left( 
\frac{V'}{V}\right)^2 \,.
\end{eqnarray} 
Recalling that $n_{{\rm T}}$ is redundant due to the consistency equation, we 
see that we now have only three observable quantities but four parameters to 
measure, namely $V$ and its first two derivatives and $\lambda$. It is clear 
therefore that a unique reconstruction is no longer possible.

This can be made explicit by redefining variables to absorb the degeneracy. 
Defining
\begin{equation}
\alpha = V \lambda^{-2/3} \;\; ; \;\; \beta = V' \lambda^{-1/2} \;\; ; \;\; 
\gamma = V'' \lambda^{-1/3} 
\label{equ18}
\,,
\end{equation}
gives a closed system for these three variables in terms of the 
observables, which can be inverted to give
\begin{eqnarray}
\alpha^3 & \simeq & \frac{25}{8} M_4^4 \, A_{{\rm T}}^2 \,;\\
\beta^2 & \simeq & \frac{25\pi}{3} M_4^2 \, A_{{\rm T}}^2 \, \frac{A_{{\rm 
T}}^2}{A_{{\rm S}}^2} \,;\\
\gamma & \simeq & 25^{2/3} \, \frac{\pi}{2} \, M_4^{2/3} 
	A_{{\rm T}}^{4/3} \left[ 4 \frac{A_{{\rm T}}^2}{A_{{\rm S}}^2} 
	+ (n-1) \right] \label{equ21} \,.
\end{eqnarray}
These resemble the usual reconstruction equations (see e.g.~Ref.~\cite{LLKCBA}). 
However the dependence on $\lambda$ is not a linear one, so its 
presence alters the functional form of the reconstructed potential and is not 
simply a scaling.

Finally, the validity of the high-energy approximation requires $V \gg 
2\lambda$, which equivalently can be written
\begin{equation}
A_{{\rm T}}^2 \gg \frac{64}{25} \, \frac{\lambda}{M_4^4} \,.
\end{equation}

That $\lambda$ cannot be determined in the high-energy limit is no surprise, 
because in that limit it appears only in the combination $M_4^2 \lambda$ which 
determines the five-dimensional Planck mass. As observations determine only 
dimensionless quantities, the overall mass scale cannot be determined and so the 
degeneracy in $M_5$ must be exact. This is precisely the same reason that one 
does not expect to determine the Planck mass from the perturbations in the usual 
four-dimensional case.

\begin{figure}[t]
\includegraphics[width=\linewidth]{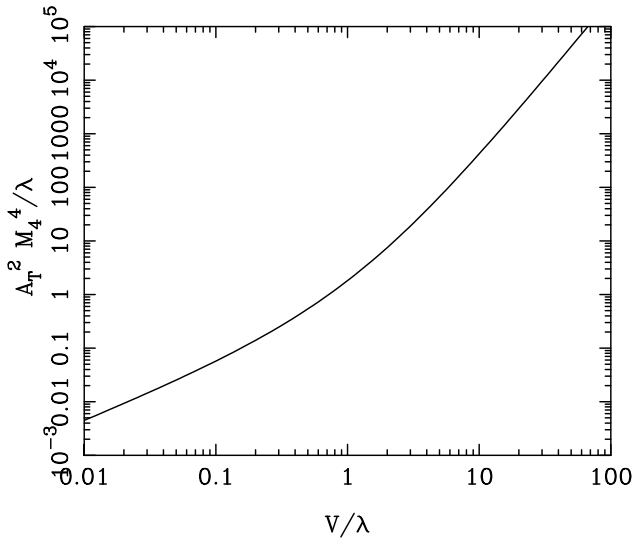}\\
\caption[fig1]{\label{fig1} Eq.~(\ref{equ2}) in dimensionless form.}
\end{figure}

\subsection{The general case}

In the general case, the perturbation spectra in the braneworld 
are given by
\ba
A_{{\rm S}}^2 & \simeq & \frac{32}{75 M_4^4} \, \frac{V}{\epsilon_{{\rm B}}} 
	\left(1+\frac{V}{2\lambda} \right) \, \left(1+\frac{V}{\lambda}
	\right) \,, \label{equ1} \\
A_{{\rm T}}^2 & \simeq & \frac{32}{75 M_4^4} \, V \,
	\left(1+\frac{V}{2\lambda}\right) G^2(V/\lambda)\,,\label{equ2}\\
%R & \simeq & \epsilon_{{\rm B}} \frac{G(V/\lambda)}{(1+V/\lambda)} 
%= -n_T/2 \,,\label{equ3}\\
n-1 & \simeq & -6 \epsilon_{{\rm B}} + 2 \eta_{{\rm B}} \label{equ4} \,,
\ea	
where $G$ is a function obtained from Eqs.~(\ref{tensamp}), (\ref{functionF}) 
and (\ref{Hubble}) by $G^2(V/\lambda) \equiv F^2(H/\mu)$, and again the 
consistency equation renders an equation for $n_{{\rm T}}$ unnecessary.
In the high-energy limit $G^2(V/\lambda) \rightarrow 3 V/2\lambda$.

The brane tension $\lambda$ can be eliminated from these equations by a set of 
redefinitions to dimensionless variables
\begin{equation}
\tilde{V} \equiv \frac{V}{\lambda} \quad ; \quad \tilde{A}_{{\rm S}}^2 \equiv
A_{{\rm S}}^2 \, \frac{M_4^4}{\lambda} \quad ; \quad \tilde{A}_{{\rm T}}^2 
\equiv A_{{\rm T}}^2 \, \frac{M_4^4}{\lambda} \,.
\end{equation}
This is sufficient to demonstrate that even in the general case the degeneracy 
is exact; one is able to reconstruct a potential from a given set of 
observations for any value of $\lambda$. The relation between $\tilde{A}_{{\rm 
T}}^2$ and $\tilde{V}$ is shown in Figure~\ref{fig1}.

We now explicitly demonstrate the effects of the braneworld on the 
reconstruction of the inflaton potential. The aim in reconstruction is to take 
measurements of the various observables, corresponding to a particular 
wavenumber $k$, and 
use these to obtain the potential and its derivatives at the scalar field value 
$\phi$ (which without loss of generality can be taken to be zero) when that 
scale crossed the Hubble radius during inflation. As 
Eqs.~(\ref{equ1})--(\ref{equ4}) are not analytically invertible, we must proceed 
using numerical inversion.

In order to obtain $V$ from a measurement of the tensors, we invert
Eq.~(\ref{equ2}) using a Newton--Raphson root-finding method.  A different value
of $V$ will be found for each choice of $\lambda$, and note from
Figure~\ref{fig1} that the function is always uniquely invertible.  In the limit
$\lambda \rightarrow \infty$ we recover the results of standard inflation, while
for small $\lambda$ the effects of the brane lead a given tensor amplitude to
correspond to a lower potential magnitude, $V$.

\begin{figure}[t]
\includegraphics[width=\linewidth]{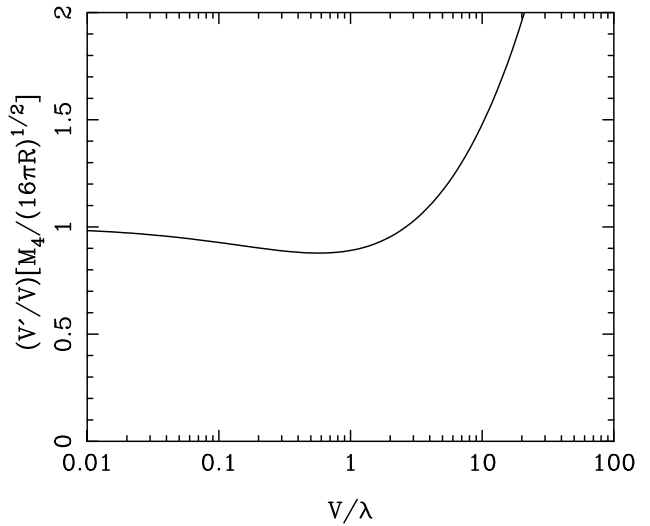}\\
\caption[fig2]{\label{fig2} The recovered slope of the inflaton potential as a 
function of 
$V/\lambda$.}
\end{figure}

To obtain the slope of the potential we invert the ratio of Eqs.~(\ref{equ2}) 
and (\ref{equ1}) to give
\be
\frac{V'}{V} = \sqrt{\frac{16 \pi R}{M_4^2}} 
	\left[\frac{1+\tilde{V}/2}{G(\tilde{V})}\right] \,,
\ee
which depends on the observable $R$ and the degenerate combination $V/\lambda$. 
Figure~\ref{fig2} shows the general relation between $V'/V$ and $V/\lambda$. 
For $V \ll \lambda$, the term in the square brackets tends to 
unity and we recover the standard inflationary result, while 
for $V \gg \lambda$ the relation approaches the high-energy limit
$V'/V \approx \sqrt{8 \pi RV/3 \lambda M_4^2}$, leading to a rapid 
increase in the obtained gradient of the inflation potential as $\lambda$ is 
reduced.

\begin{figure}
\includegraphics[width=\linewidth]{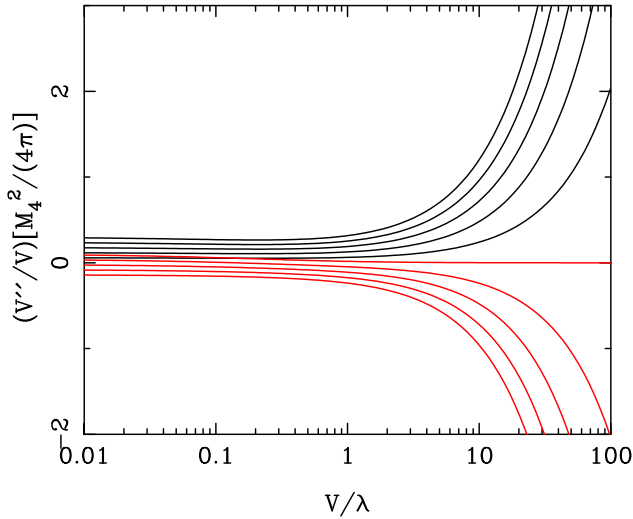}\\
\caption[fig3]{\label{fig3} The recovered curvature of the inflaton potential in 
the 
braneworld scenario as 
a function $V/\lambda$. The upper set of five lines are for  $R=0.01$ to 
$0.05$, with $n-1=0$, while the lower set are from the same range of $R$ with 
$n-1=-0.2$.}
\end{figure}

The second derivative of $V$ can be obtained from 
\be
\frac{V''}{V} = \frac{4 \pi}{M_4^2}  \left(1+\frac{\tilde{V}}{2} \right) \, 
	\left[  6 R \, \frac{1+\tilde{V}}{G^2(\tilde{V})} + (n-1) \right] \,,
\ee
which is a function of the observables $R$ and $n-1$, plus the degenerate 
combination $V/\lambda$. Figure~\ref{fig3} shows the recovered curvature
of the potential as a function of $V/\lambda$ for a 
range of values of $R$ and $n-1$. For $V \gg \lambda$ the 
magnitude of the curvature of the potential increases and asymptotes to 
Eq.~(\ref{equ21}). 

Finally we demonstrate these degeneracies for an example set of observables.  We
choose our observables to be $A_T^2=4 \times 10^{-12}$, $R=0.01$, and
$n-1=-0.05$; these numbers correspond roughly to the predictions from a quartic
potential and are consistent with current observations \cite{WTZ} (the ratio of
contributions to the large-angle microwave anisotropies is about $4\pi R$ in our
conventions).  We reconstruct potentials for choices of $\lambda$ evenly spaced 
logarithmically in the range from $10^{-15} \, M_4^4$ to $10^{-9} \, M_4^4$.  In 
each case, we plot only the portion of the potential accessible to
observations; the relation between $\Delta \phi$ and the range of scales probed 
by observations depends on $\lambda$ and is computed via Eq.~(\ref{efolds}) as
\be
\frac{\Delta \phi/M_4}{\Delta \ln k}  = \sqrt{\frac{R}{4 \pi}} \; 
\frac{1}{G(\tilde{V})} \,.
\ee
We take the Planck satellite as our guideline, which will have 
$\Delta \ln k \simeq 3.5$ on either side of the central point.  

\begin{figure}[t]
\includegraphics[width=\linewidth]{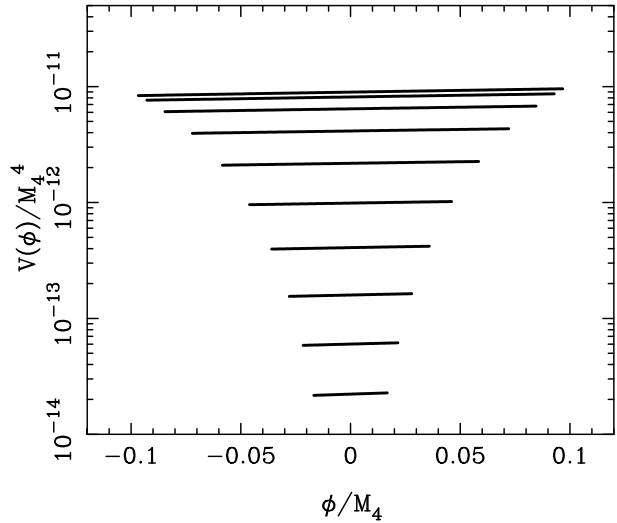}\\
\caption[fig4]{\label{fig4} Different potentials as obtained for the different 
choices of $\lambda$ described in the text, with the highest curves 
corresponding to the highest $\lambda$. Einstein gravity is recovered as 
$\lambda \rightarrow \infty$. Each potential generates the same initial 
spectra.}
\end{figure}

Figure~\ref{fig4} shows a set of reconstructed model potentials
for the different assumed values of $\lambda$, each of which reproduces our 
model
observations. The ratio $V/\lambda$ obtained ranges from $0.01$ to
$22$. For $V \ll \lambda$ the reconstructed potential is nearly 
independent of $\lambda$, closely approximating the Einstein gravity result. As 
$\lambda$ is decreased, the magnitude of 
the potential begins to decrease while its gradient steepens; at the same time 
the amount of potential constrained shrinks as the extra friction leads to 
slower rolling of the field. 

To end, we mention that as well as reproducing the correct perturbations, a
viable potential must be able to support enough subsequent inflation to stretch
those perturbations to the observable scales.  If the recovered potential
develops a minimum or goes negative within the constrained range, our approach
will have broken down, and refinement becomes necessary (going beyond the
quadratic potential approximation and/or slow-roll) to test whether there is
still a viable potential. However the approximate condition for the 
reconstruction to break down, $\Delta \phi \gtrsim |V'/V''|$, does not change 
significantly as $\lambda$ is decreased, because in the high-energy limit 
$\Delta \phi$ reduces at the same rate as $V'/V''$ (for fixed values of the 
observables). Hence if a viable potential exists in the $\lambda \rightarrow 
\infty$ limit, it is unlikely that the problem will become ill-defined for low 
values of $\lambda$.

\section{Conclusions}

One of the most anticipated results of forthcoming high-accuracy CMB experiments
is the probing of the physics of inflation, and in particular empirically
reconstructing the form of the inflaton potential.  However, it is important to
be aware of the possible degeneracies that may arise.  To date attention has
been focussed on degeneracies between initial perturbation parameters and
cosmological parameters such as reionization, suggesting that combinations of
observations (for example CMB polarization as well as temperature, or completely
different types of observation) are required to lift these degeneracies.

For early Universe cosmologists, more worrying are degeneracies that arise in
predictions for the initial perturbations, which represent a fundamental
limitation to the constraints we can extract and which are not broken by 
polarization.  In this paper we have described
how such a degeneracy arises in a braneworld scenario based on the
Randall--Sundrum type-II model.  We have shown that the unique reconstruction of
the potential from scalar and tensor perturbation spectra in this scenario is no
longer possible, with a different possible potential arising for each
choice of brane tension.  Accordingly, observations of the perturbation spectra
cannot distinguish between the braneworld and standard inflation.  It would be
interesting to know if this is unique to the simplest braneworld scenario, or if 
it remains true in other versions. It would also be interesting to know if this 
result persists at higher order in the slow-roll expansion for the 
perturbations.

We end by stressing that our results refer to the initial perturbation spectra.
Whether or not there might be significant braneworld effects on the subsequent 
evolution of
the perturbations is presently unknown and is likely to be model dependent; for
example in general the short-scale perturbation behaviour on the brane can be
influenced by bulk perturbations which cannot be predicted on the brane (see
Ref.~\cite{Maartens} for an overview).  It may well be that the braneworld might
manifest itself through such effects.  If, however, the perturbation evolution
turns out to be unaffected (for example if inflation is successful in diluting
the effect of bulk perturbations), then finding observable traces of the
braneworld in the low-energy universe may not be easy.

%%%%%%%%%%%%%%%%%%%%%%%%%%%%%%%%%%%%%%%%%%%%%%%%%%%%%%%%%%%%%%%%%%%%%%%%
\begin{acknowledgments}
A.R.L.~was supported in part by the Leverhulme Trust. A.N.T.~was supported by a 
PPARC Advanced Fellowship, and thanks the University of Sussex Astronomy 
Centre, where this work began, for its hospitality. We thank James Lidsey and 
David Wands for discussions.
\end{acknowledgments}

%%%%%%%%%%%%%%%%%%%%%%%%%%%%%%%%%%%%%%%%%%%%%%%%%%%%%%%%%%%%%%%%%%%%%%%%
 
%%%%%%%%%%%%%%%%%%%%%%%%%%%%%%%%%%%%%%%%%%%%%%%%%%%%%%%%%%%%%%%%%%%%%%%

\begin{thebibliography}{}
\bibitem{LL} A. R. Liddle and D. H. Lyth, {\em Cosmological Inflation
	and Large-Scale Structure}, Cambridge University Press,
	Cambridge, 2000.
\bibitem{CKLL} E. J. Copeland, E. W. Kolb, A. R. Liddle, and J. E. Lidsey,
	Phys. Rev. D{\bf 48}, 2529 (1993) [hep-ph/9303288].
\bibitem{LLKCBA} J. E. Lidsey, A. R. Liddle, E. W. Kolb, E. J. Copeland,
	T. Barreiro, and M. Abney, Rev. Mod. Phys. {\bf 69}, 373 (1997)
	[astro-ph/9508078].
\bibitem{bin} P. Bin\'etruy, C. Deffayet, and D. Langlois, Nucl. Phys. 
	{\bf B565} (2000) 269 [hep-th/9905012]; P. Bin\'etruy, C. Deffayet, 
	U. Ellwanger, and D. Langlois, Phys. Lett. B{\bf 477}, 285 (2000)
	[hep-th/9910219]. 
\bibitem{shiro} T. Shiromizu, K. I. Maeda, and M. Sasaki, Phys. 
	Rev. D{\bf 62}, 024012 (2000) [gr-qc/9910076]. 
\bibitem{RSII} L. Randall and R. Sundrum, Phys. Rev. Lett. {\bf 83}, 4690 
	(1999) [hep-th/9906064].
\bibitem{MWBH} R. Maartens, D. Wands, B. A. Bassett, and I. P. C. Heard,
	Phys. Rev. D{\bf 62}, 041301 (2000) [hep-ph/9912464].
\bibitem{LMW} D. Langlois, R. Maartens, and D. Wands, Phys. Lett B{\bf 489}, 
	259 (2000) [hep-th/0006007].
\bibitem{HL} G. Huey and J. E. Lidsey, astro-ph/0104006.
\bibitem{TB00} A. N. Taylor and A. Berera, Phys. Rev. D{\bf 62}, 083517 (2000) 
	[astro-ph/0006077].
\bibitem{WTZ} X. Wang, M. Tegmark, and M. Zaldarriaga, astro-ph/0105091.
\bibitem{Maartens} R. Maartens, gr-qc/0101059.
\end{thebibliography}
\end{document}